\documentclass[svgnames]{article}
\usepackage[
a4paper,includeheadfoot,%
headheight=12pt,
headsep=10pt,%
footskip=5pt,
marginparwidth=38pt]{geometry}
\usepackage{fancyhdr}
\usepackage{lshtmcolours}
\usepackage{setspace}
\usepackage{amssymb}
\usepackage{extarrows}
\usepackage{hyperref}
\usepackage[numbers,compress]{natbib}
\usepackage{doi}
\hypersetup{colorlinks=true,allcolors=ForestGreen}
\usepackage{tabularx}
\usepackage{booktabs}
\usepackage{literate}
\usepackage{subcaption}
\usepackage{yfonts}
\usepackage{fancybox}
\usepackage{enumitem}
\usepackage{todonotes}
\usepackage{pgfplots}
\usepackage{immune}
\usepackage{envelope}
\usepackage{wrapfig}
\usepackage{authblk}
\usetikzlibrary{external}

\begin{document}
\renewcommand\Affilfont{\itshape\footnotesize}
\title{Compositional modelling of immune response and virus transmission dynamics}

\author[1,2]{William Waites}
\author[3]{Matteo Cavaliere}
\author[2,4]{Vincent Danos}
\author[5]{Ruchira Datta}
\author[1]{Rosalind M. Eggo}
\author[6]{Timothy B. Hallett}
\author[7]{David Manheim}
\author[8,9]{Jasmina Panovska-Griffiths}
\author[1]{Timothy W. Russell}
\author[10]{Veronika I. Zarnitsyna}

\affil[1]{Centre for Mathematical Modelling of Infectious Disease, London School of Hygiene and Tropical Medicine}
\affil[2]{School of Informatics, University of Edinburgh}
\affil[3]{Department of Computing \& Mathematics, Manchester Metropolitan University}
\affil[4]{D\'{e}partement d'Informatique, \'{E}cole Normale Sup\'{e}rieure, Paris}
\affil[5]{Datta Enterprises LLC}
\affil[6]{MRC Centre for Global Infectious Disease Analysis, Imperial College London}
\affil[7]{Technion, Israel Institute of Technology}
\affil[8]{The Big Data Istitute, Nuffield Department of Medicine, University of Oxford}
\affil[9]{The Queen's College, University of Oxford}
\affil[10]{Department of Microbiology and Immunology, Emory University School of Medicine}

\maketitle


\begin{abstract}
Transmission models for infectious diseases are typically formulated in terms of dynamics between individuals or groups with processes such as disease progression or recovery for each individual captured phenomenologically, without reference to underlying biological processes. Furthermore, the construction of these models is often monolithic: they don't allow one to readily modify the processes involved or include the new ones, or to combine models at different scales.  We show how to construct a simple model of immune response to a respiratory virus and a model of transmission using an easily modifiable set of rules allowing further refining and merging the two models together. The immune response model reproduces the expected response curve of PCR testing for COVID-19 and implies a long-tailed distribution of infectiousness reflective of individual heterogeneity. This immune response model, when combined with a transmission model, reproduces the previously reported shift in the population distribution of viral loads along an epidemic trajectory. 
\end{abstract}

\fancyhead{}
\fancyfoot{}
\fancyhead[RE,RO]{\thepage}
\renewcommand{\headrulewidth}{0pt}
\pagestyle{fancy}
\thispagestyle{empty}

\section{Introduction}
Our starting point in this paper is work by Hellewell and
colleagues~\cite{hellewell_estimating_2021} where they fit a descriptive
statistical model to empirical data~\cite{houlihan_pandemic_2020} describing
cycle thresold (Ct) values for PCR tests of healthcare workers in England.
Rather than describing the data, we aim to get at the mechanism behind it.
We begin by asking: what is the simplest, biologically reasonable, process that
is sufficient to produce such data?
Our answer to this question also produces insights into a possible biological
basis for over-dispersion of infections~\cite{endo_estimating_2020} and, when
an infection transmission process is added, the observed shifts in distribution
of Ct values during the rising and falling of an
epidemic~\cite{hay_estimating_2021}.

The method that we use builds upon previous theoretical
work in applying techniques first developed for
molecular biology~\cite{danos_formal_2004,boutillier_kappa_2018} to
epidemics~\cite{waites_rule-based_2021}.
These techniques have two important features: \emph{modularity} and
\emph{compositionality}.
Modularity means that it is possible to create self-contained models, in this
case of immune response and transmission that can be individually calibrated and
studied.
Compositionality means that we can combine these models to create larger ones.
This is important for several reasons.
Some operations, particularly calibration or fitting to data, are both
computationally expensive and critical for real-world applications.
We would like, so far as possible, to do these operations once and reuse them.

Beyond the immediate subject-matter of this paper, epidemic trajectories
strongly influence and are influenced by processes at several spatial and
temporal scales: biology and physiology of the hosts, behaviour and individual
choices, policy choices and decision-making, and economic environments to name a
few. 
All of these topics require sophisticated modelling efforts in their own right
and a substantial amount of domain expertise.
Any strategy for understanding the interplay of these processes that relies on a
monolithic approach for modelling is unlikely to be feasible.
A modular approach that breaks the large system down into components
individually modelled at appropriate scale~\cite{zarnitsyna_advancing_2021} and
then composes them into a model of the whole is much more likely to be
effective because the individual components can be small enough to be tractable
and enable domain experts to focus their expertise on them.
It is equally important that the composition of such models are
well-defined if we are to understand the properties of the combined model.
We use a method of doing this that builds on a solid foundation of abstract
mathematics and category theory~\cite{danos_formal_2004,behr_stochastic_2016,behr_nicolas_rule_2020}
to ensure that this is the case, as well as properly representing the biological and epidemiological processes.

Multi-scale models of within- and between-host transmission
exist and are recognised as an important line of
research~\cite{handel_crossing_2015,garabed_multi-scale_2020} and
attention has been given to the possibility of composite models for this
purpose~\cite{garira_complete_2017}.
Nevertheless, the bulk of infectious disease models~\cite{abbott_epinow2_2020,ferguson_covid-19_2020,kerr_covasim_2021} are not formulated in ways
amenable to composition, perhaps because methods for defining interfaces between
models and the interaction of time-scales and concepts of discrete and
continuous time remain poorly understood.
This remains an important challenge, however progress can be made.
There is a longstanding recognition of the need for
modular composable models and there are well-developed efforts in this direction in synthetic and systems biology~\cite{hucka_systems_2003,le_novere_biomodels_2006,lloyd_cellml_2008,smolke_building_2009,chelliah_biomodels_2013,clerx_cellml_2020,keating_sbml_2020,blinov_practical_2021}.

The remainder of this paper is structured as follows. We first briefly introduce
rule-based models in general. We then give our model of adaptive immune response
and show the implications for the expected distribution of viral load in terms
of time since infection that follows from it. Next, we give the model of
transmission and how it is coupled to the immune model and show how the
distribution of viral load can be expected to vary according to an epidemic
trajectory. Finally we discuss some limitations and future
research directions.

\section{Rule-based models}
\label{sec:rules}

The model that we present here, and its constituent sub-models, are formulated as
rules. Informally, a rule is the way states of the model change.
A rule is a structure with a left-hand side $\mathcal{L}$, a right-hand side
$\mathcal{R}$, and a rate, $k$.
$\mathcal{L}$ is interpreted as a pattern that matches some part of the system, for example, people with a given characteristic or state or even relationships.
$\mathcal{R}$ is an instruction for how the configuration should be changed when
that pattern matches;
the action of a rule is to rewrite the configuration of the system so that it evolves over time.
In a single simulation step, a single match, or embedding, is chosen from all possible
embeddings $\mathcal{E}(\mathcal{L})$ in the graph representing the state of the system, and this embedding is replaced according
to the instruction. 
Thus, a rule encodes many possible transitions (one for
each possible embedding) in a continuous time Markov chain (CTMC) where the state-space is all possible graphs.
This Markov chain can be simulated with the usual method where the propensity of
a rule to operate is given by the number of embeddings, and the rate,
$p = k|\mathcal{E}(\mathcal{L})|$.
It is possible to sample trajectories from this CTMC exactly without materialising the entire state-space or employing moment-closure techniques~\cite{boutillier_kappa_2018}.
It is also possible to systematically produce ordinary differential equations for the
moments of the CTMC, but they are rarely feasible to use in practice for any but the
simplest\footnote{A typical stratified infectious disease model might have, when rendered as a system of ODEs, at most a hundred dimensions and perhaps a thousand terms. The simple rule-based model that we present here, though finite, corresponds to a system of ODEs with nearly ten thousand dimensions and nearly a hundred million terms.} models because of the high dimensionality involved.

The algebra of rules that governs how rules compose is an advanced topic~\cite{behr_stochastic_2016}, but it is easy to obtain an intuition about how it works with a version of Gillespie's algorithm~\cite{gillespie_exact_1977}.
Suppose that we have two rules, $r_1$ and $r_2$.
To perform one simulation step for a model with only one rule, we choose one match uniformly at random of the left-hand side of the rule in the system state, replace it with the right-hand side and advance time by an exponentially distributed amount with rate given by the rule's propensity.
Composing the two rules, we perform a step by first choosing which rule to use with probability proportional to their propensities.
Having chosen a rule, we then choose a match uniformly at random, do the replacement, and advance time as above with a rate of the total propensity of all rules.
This is exactly analogous to simulating a single chemical reaction, except that the ``match'' and ``replace'' operations are more complex because they involve searching in and manipulating a graph as opposed to counts of chemical species.
Fortunately, we have software that can do this efficiently~\cite{boutillier_kappa_2018}.

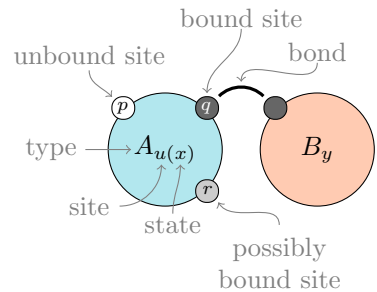
\begin{wrapfigure}{r}{0.4\textwidth}
  \begin{center}
    \begin{tikzpicture}
      \node[draw,fill=S,circle,minimum size=1.5cm] (a) at (0,0) { $A_{u(x)}$ };
      \node[draw,rounded corners,fill=white,minimum size=0.3cm,above left=-0.2cm of a] (unbound) {};
      \node[draw,rounded corners,fill=black!60,minimum size=0.3cm,above right=-0.2cm of a] (bound) {};
      \node[draw,rounded corners,fill=black!20,minimum size=0.3cm,below right=-0.2cm of a] (maybe) {};
      \node[above left=-0.27cm of a] { $_p$ };
      \node[above right=-0.27cm of a] { \color{white} $_q$ };
      \node[below right=-0.24cm of a] { $_r$ };
      
      \node[draw,fill=I,circle,minimum size=1.5cm] (b) at (2,0) { $B_y$ };
      \node[draw,rounded corners,fill=black!60,minimum size=0.3cm,above left=-0.2cm of b] (bound2) {};
      \path[draw,very thick] (bound) to[out=45,in=135] (bound2);
      
      \path[draw,->,black!50] (-0.6,-0.75) to[out=0,in=270] (-0.05,-0.2);
      \node[black!50] at (-1,-0.75) { site };
      \path[draw,->,black!50] (0.1,-0.85) to[out=90,in=270] (0.2,-0.2);
      \node[black!50] at (0.1,-1) { state };
      \path[draw,->,black!50] (-1,1.05) to[out=270,in=135] (-0.7,0.7);
      \node[black!50] at (-1,1.25) { unbound site };
      \path[draw,->,black!50] (1,1.6) to[out=270,in=90] (0.55,0.75);
      \node[black!50] at (1,1.75) { bound site };
      \path[draw,->,black!50] (2,1.15) to[out=270,in=90] (1,0.9);
      \node[black!50] at (2,1.3) { bond };
      \path[draw,->,black!50] (-1.05,0) to[out=0,in=180] (-0.45,0);
      \node[black!50] at (-1.5,0) { type };
      \path[draw,->,black!50] (1.5,-1) to[out=90,in=300] (0.7,-0.7);
      \node[black!50] at (1.5,-1.5) { \begin{minipage}{2cm}\centering possibly\\bound site\end{minipage} };
    \end{tikzpicture}
  \end{center}
  \caption{Illustration of an agent pattern with features identified.}
  \label{fig:agent-pattern}
\end{wrapfigure}
In the $\kappa$ language~\cite{danos_formal_2004,boutillier_kappa_2018}, the
configuration of the system is described as a site-graph where vertices are
called \emph{agents} (by analogy with the reagents in chemistry, not to be
confused with the agents of agent-based models, though some parallels exist), that have \emph{sites} which
can have internal \emph{state} and edges are \emph{bonds} between sites of
different agents.
The biochemical heritage of the formalism is evident from this nomenclature.
Figure~\ref{fig:agent-pattern} shows an example agent pattern (e.g. the
left-hand side of a rule) pictorially.
It consists of two agents, $A$ and $B$. Both have several binding sites which
can be in a unbound ($p$) or bound ($q$) state. For patterns, we may not care
whether a particular site is bound or not, normally such sites would not be
mentioned in a pattern, but when we wish to explicitly depict it, we do this
with a half-shaded site ($r$).
$A$ has a site $u$ with internal state $x$.
Where there is no risk of confusion, we may drop the name of the site, as with
$B$ which has a nameless binding site and a site with state $y$.
For a thorough explanation of this arrangement in practice in its original
setting, the reader may wish to consult Boutillier et al.'s original
article~\cite{boutillier_kappa_2018}, as well as our tutorial with
examples on the application to infectious
diseases~\cite{waites_rule-based_2021}.

We will use several agent types in the models given below. Agents of type $P$
represent people -- chosen rather than $I$ for individual to avoid confusion
with individuals in an infectious state as appears in many susceptible,
infectious, removed (SIR) style models. Elements of immune response are
represented by agents bound to a person, $B$ for B-cells, $A$ for antibody
populations. The viral population within a host is represented as an agent of type
$V$. Diagnostic tests of various types are explicitly represented as agents of
type $T$.

\section{Adaptive Immune Response}
\label{sec:immune}


While we use a simplified model here, the general approach can be extended to more complex models of immune response with relatively little effort. 
We give the model in mathematical form, with explanatory narrative.
A methodological observation is that, though we use shape and colour for visual
interest, there is an exact correspondence between what is depicted and the
machine-readable computer code used for simulation reproduced in
Appendix~\ref{sec:code}.

\subsection{Viral load}
\label{sec:viral-load}
The entire within-host response is driven by replication of the virus.
We do not represent individual virions in this model and simply track the size
of the virus population.
We track this with a counter that represents the logarithm of the population
size.
Replication is captured simply by incrementing this counter,
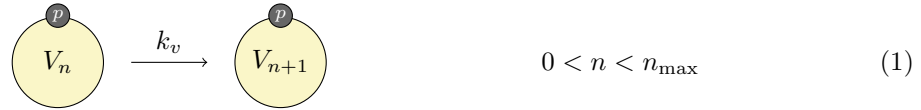
\begin{align}
  \replicationl &\arate{k_v} \replicationr & 0 < n < n_{\text{max}}
\end{align}
up to some maximum.
In principle, uncontrolled replication should proceed up to the carrying capacity of the
host.
For simplicity, we represent this process as simple exponential growth with a
maximum limit rather than logistic growth.

\subsection{Initial activation}
The peak of viral replication is usually controlled by the innate response together with target cell limitation and further clearance of the virus is controlled by CD8 T-cells and antibodies~\cite{baccam_kinetics_2006,saenz_dynamics_2010,miao_quantifying_2010}.
In this simplified model, we do not represent the innate response, and we consider notional antibodies as generalised effectors providing the only mechanism for clearance of viral particles.
We begin with the
activation of B-cells in an infected individual bound to a virus population,
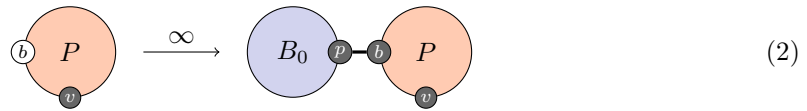
\begin{equation}
  \tcelll \arate{\infty} \tcellr
\end{equation}
On the left-hand side, the individual has no activated B-cells, and on the
right-hand side, they acquire a population of activated B-cells.
The subscript on $B$ indicates that all B-cells in this population are naive:
they have very low affinity for viral proteins.
This process is interpreted as \emph{allocation of a population} of B-cells and
not production of the cells themselves.
This is somewhat of a computational fiction and is an artefact of the
simplifying choice as with the virus population to not track individual cells,
but only their count (or in this case, their affinity).
The process therefore proceeds immediately, at infinite rate.

The function of B-cells is to produce antibodies. Again, as with the virus population,
we do not track individual antibody proteins, we simply track their number,
again logarithmically.
The second process in the activation of immune response is,
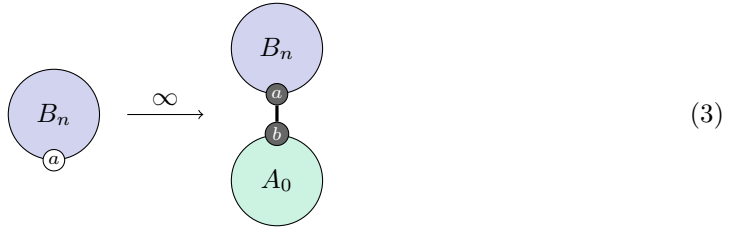
\begin{equation}
  \bcelll \arate{\infty} \bcellr
\end{equation}
As with the activation of B-cells, this process is interpreted as
allocation of a population of antibodies, not the production of
antibodies themselves.
Antibody production is a separate rule (Equation~\ref{eq:antibody}.
As with the allocation of a B-cell population, the allocation of an antibody population proceeds immediately.

\begin{wrapfigure}{r}{0.3\textwidth}
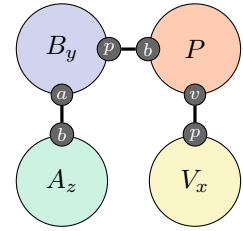

  \begin{center}
    \fullimmune{x}{y}{z}
  \end{center}
  \caption{Fully active immune response with viral load $x$, B-cell affinity
    $y$, and antibody count $z$.}
  \label{fig:fullimmune}
\end{wrapfigure}
Once the immune response has been activated, the configuration of the
within-host state for an individual is as depicted in
Figure~\ref{fig:fullimmune}. The individual is bound to a virus population of a
certain size, $x$ (the process by which this happens is exogenous to the immune
model),
The individual is also bound to a B-cell population with average affinity $y$,
and these B-cells are bound to an associated antibody population with count $z$.

\subsection{Immune dynamics}
\label{sec:immunedynamics}
Now we can describe the dynamic processes of the adaptive immune response.
They begin with affinity maturation.
The process of affinity maturation is only partly understood.
Virus-specific B-cells, initially with low affinity, are recruited into special structures called germinal centers (GCs), where their affinity maturation occurs. In GCs, B-cells get stimulated through their B cell receptors by the antigen and undergo the rounds of proliferation with somatic hypermutation (SHM) of their immunoglobulin genes. The cycles of SHM coupled to selection for antigen binding lead to the generation of B-cells of higher affinity.
During the affinity maturation process, a fraction of B-cells further differentiate into plasma cells that begin to secrete virus-specific antibodies~\cite{krautler_differentiation_2017}.
Somatic hypermutation continues, and the
antibodies produced by the B-cells are iteratively refined and better tuned
for binding to the virus.
We elide the detail of this fascinating process and treat it phenomenologically at a higher level of abstraction.
We reason that in a well-mixed environment, the chance of a B-cell encountering
a compatible virion is proportional to some function of the virus population and
write,
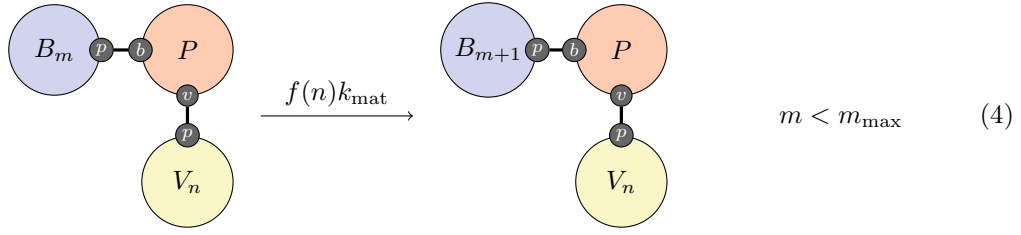
\vspace{0.25\baselineskip}
\begin{align}
  \affinityl &\arate[2cm]{f(n)k_{\text{mat}}} \affinityr & m < m_{\text{max}}
\end{align}
For simplicity, we take $f(n) = n$ and obtain $k_{\text{mat}}$ by calibration.

By similar reasoning, antibodies are produced at a rate that we take to be
proportional to the affinity,
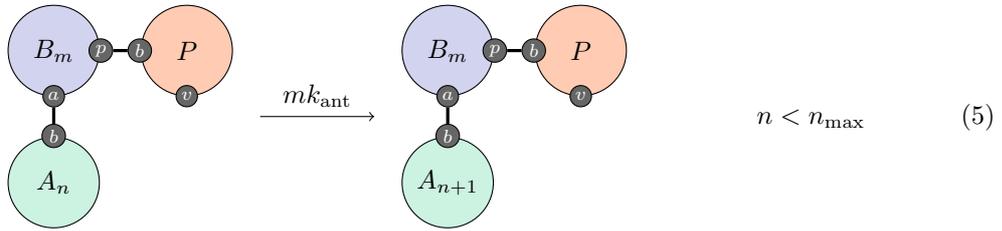
\vspace{0.25\baselineskip}
\begin{align}
  \label{eq:antibody}
  \antibodyl &\arate[1.5cm]{mk_{\text{ant}}} \antibodyr & n < n_{\text{max}}
\end{align}
and likewise obtain $k_{\text{ant}}$ by calibration.
Note that the $P$ agent appears in this rule with an explicitly bound virus
population. This is because virus-specific B-cells populations arise and are activated only in the presence of viral
proteins; if there are no such proteins, negligible amounts of antibodies (none in our model) are produced.

The action of antibodies is to disable or neutralise virions.
We follow a similar pattern and suppose that this happens at a rate proportional
to the antibody population,
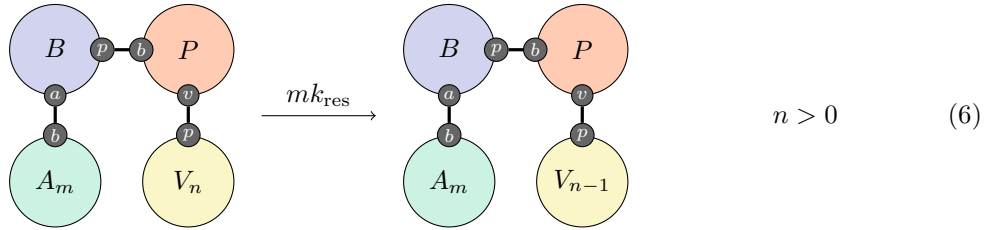
\vspace{0.25\baselineskip}
\begin{align}
  \neutrall &\arate[1.5cm]{mk_{\text{res}}} \neutralr & n > 0
\end{align}
with the constant of proportionality $k_{\text{res}}$ again obtained by
calibration.

Finally, recovery is said to happen when the virus population drops sufficiently
low,
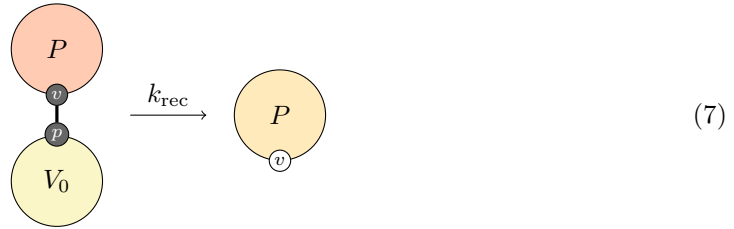
\begin{equation}
  \recoveryl \arate{k_{\text{rec}}} \recoveryr
\end{equation}
Though we do not need to specify it in this rule, the recovering
individual will still have a population of B-cells with some affinity and its
associated antibody population.
We do not have a rule that severs those edges, nor do we have a rule that can
decrease affinity.
This provides a mechanism for immune memory. If a previously infected individual
is exposed to the virus again, they may have a population of antibodies that can
immediately dispatch it and if they do not, their B-cells can produce them.

Finally, we include a simple rule very similar to the above for waning, where the
antibody population decreases slowly when the individual does not have a virus
population,
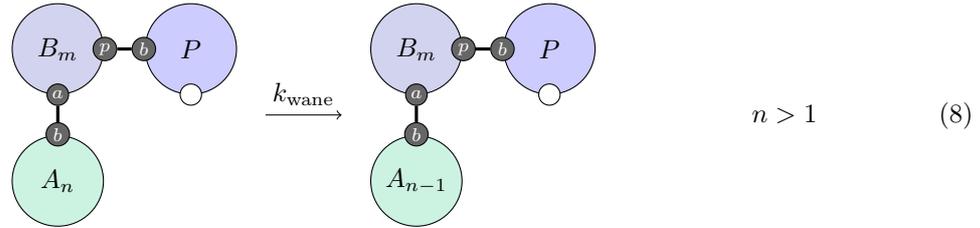
\begin{align}
  \waningl &\arate{k_\text{wane}} \waningr & n > 1
\end{align}

\subsection{Connection to compartments}
\label{sec:observables}
In epidemiological models it is traditional to subdivide the population into
susceptible individuals, removed individuals, and one or more compartments of
individuals in different states of incubation or infectiousness.
Even with individual- or agent-based models, this classification of individuals is commonly used.
There is a natural way to make the connection to these concepts simply by
counting configurations of individuals and their immune response.
This is done by defining \emph{observables} that count the number of embeddings of a pattern similar to the operation of the left-hand side of a rule,
\vspace{0.25\baselineskip}
\begin{align}
  S &= \left|\;\tikz[baseline]{
      \node[agent,fill=S] (a) at (0,0.1) {P};
      \node[site,fill=white,left=-0.15cm of a] (pv) { $\scriptstyle b$ };
      \node[site,fill=white,below=-0.15cm of a] (pv) { $\scriptstyle v$ };
      }\;\right|
  &
  I &= \left|\;\tikz[baseline]{
      \node[agent,fill=I] (a) at (0,0.1) {P};
      \node[site,fill=black!60,below=-0.15cm of a] (pv) { \color{white}$\scriptstyle v$ };
      }\;\right|
  &
  R &= \left|\;\tikz[baseline]{
      \node[agent,fill=R] (a) at (0,0.1) {P};
      \node[site,fill=black!60,left=-0.15cm of a] (pv) { \color{white}$\scriptstyle b$ };
      \node[site,fill=white,below=-0.15cm of a] (pv) { $\scriptstyle v$ };
      }\;\right|
\end{align}
\vspace{0.25\baselineskip}
where the vertical bars are used in the sense of cardinality of a set -- the set
of embeddings induced by the patterns.
A susceptible individual is one who has no virus population and no established
immune memory.
An infected or infectious individual has a virus population regardless of the state of their immune response and a recovered or removed individual has no virus population but does have
established immune memory.
These correspondences of course only give a coarse picture of the expected
dynamics: they do not consider the robustness of the immune memory or the degree
of infectiousness.
This coarseness is the reason that, if we were to work at the level of
compartments (i.e. high-level disease progression states whether in a compartmental model or otherwise), we would be required to formulate processes in terms of complicated distributions
chosen either by hypothesis or empirically.
Here, we do not need to do that because we have access to a fine-grained account
of adaptive immune response.
It is nevertheless useful to have a view of the dynamics of this model that is
comparable to the more common representation.

\subsection{Empirical validation}
\label{sec:empirical}
It is easy to see that viral load -- the quantity of virus carried by a host --
must vary over the course of the disease.
At the time immediately prior to infection, there is no virus present.
At some future time after the infection has been cleared, there is likewise no
virus present.
At some time in between there must be some non-zero amount of virus present
because that is the meaning of infection.
Therefore, viral load must have been increasing for some of that time, and
decreasing for some of that time and it must have had a maximum for
some time between these phases.
Whilst there is no \emph{a priori} guarantee that there must be only one
occurrence of each of these phases, the simplest behaviour would be a single
increasing phase, a peak, and a decreasing phase.
This is precisely what we observe empirically through assays that are
sensitive to the presence of virus itself (e.g. PCR) or viral proteins (e.g.
antigen)~\cite{to_temporal_2020,walsh_sars-cov-2_2020,cevik_virology_2020,hellewell_estimating_2021}.
Furthermore, because the various immune processes that eventually suppress the
population of virions take time to develop, any increase in antibody products
must lag behind an initial increase in viral load, again, precisely what is
observed with antibody assays~\cite{to_temporal_2020,walsh_sars-cov-2_2020,cevik_virology_2020}.

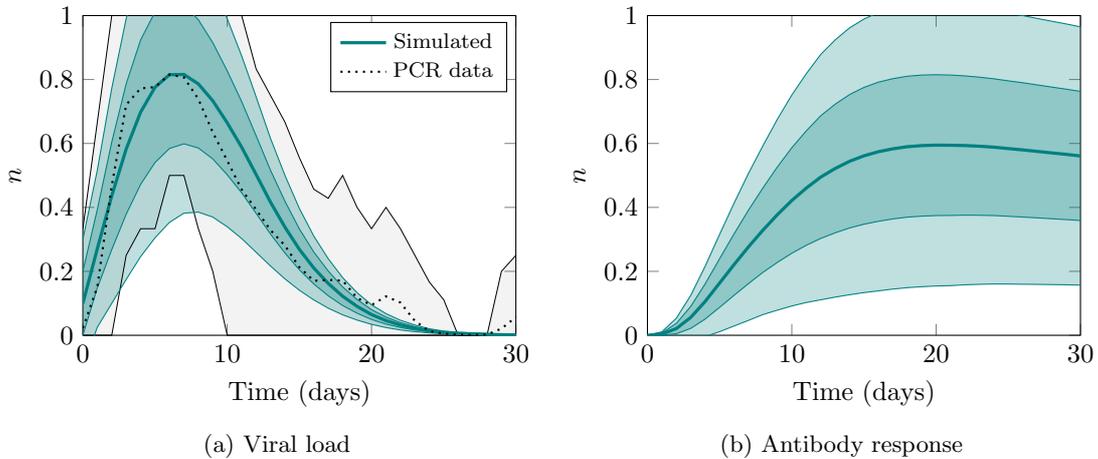
\begin{figure}
  
  \begin{subfigure}{0.495\textwidth}
    \tikzsetnextfilename{fig1a-immune-response}
    \begin{tikzpicture}
      \begin{axis}[
        cycle list name=exotic,
        width=\textwidth, height=0.8\textwidth,
        xmin=0, xmax=30, ymin=0, ymax=1,
        xlabel={Time (days)}, ylabel={$n$},
        no markers
        ]
        \addplot[forget plot,very thin,name path=etop] table[col sep=comma,x={x}, y={top}] {empirical_samples_pcr_detection.csv};
        \addplot[forget plot,very thin,name path=ebot] table[col sep=comma,x={x}, y={bottom}] {empirical_samples_pcr_detection.csv};
        \addplot[forget plot,opacity=0.05,draw=none] fill between[of=ebot and etop];
        \envelope{V}{Vstd}{immune-response.tsv};\addlegendentry{\footnotesize Simulated};
        \addplot[thick,dotted] table[col sep=comma,x={x}, y={mean}] {empirical_samples_pcr_detection.csv};\addlegendentry{\footnotesize PCR data}
      \end{axis}
    \end{tikzpicture}
    \caption{Viral load}
    \label{fig:viral_load}
  \end{subfigure}
  \hfill
  \begin{subfigure}{0.495\textwidth}
    \tikzsetnextfilename{fig1b-immune-response}
    \begin{tikzpicture}
      \begin{axis}[
        cycle list name=exotic,
        width=\textwidth, height=0.8\textwidth,
        xmin=0, xmax=30, ymin=0, ymax=1,
        xlabel={Time (days)}, ylabel={$n$},
        no markers
        ]
        \envelope{A}{Astd}{immune-response.tsv};
      \end{axis}
    \end{tikzpicture}
    \caption{Antibody response}
    \label{fig:antibody_response}
  \end{subfigure}
  \caption{Within-host immune dynamics sampled from a population of 10,000. The
    system is fit to PCR response data response as reported by Hellewell et
    al.~\cite{hellewell_estimating_2021}.
    Underlying processes of virus replication, affinity maturation, antibody
    production and virus neutralisation reproduce the characteristically
    asymmetric viral load curve (\ref{fig:viral_load}) and an antibody response
    curve (\ref{fig:antibody_response}) that lags viral load. The vertical axis
    in both figures is measured in arbitrary logarithmic units.}
  \label{fig:immune-response}
\end{figure}
The adaptive immune response model reproduces the above observations from the
underlying processes.
We calibrated the model with the Approximate Bayesian Computation (ABC) method
using a Root Mean Square (RMS) distance relative to the mean PCR test response
for SARS-CoV-2 infection as reported by Hellewell et
al.~\cite{hellewell_estimating_2021} based on data from a study of health
workers in England~\cite{houlihan_pandemic_2020}.
The fitting procedure is computationally expensive, however it need only be performed once: the calibrated immune model stands on its own and for many uses it can be combined with other models without the necessity to repeat this expensive procedure.
The resulting viral load and antibody response curves for a population of 10,000
individuals are shown in Figure~\ref{fig:immune-response} and the reference data
together with its 95\% credible interval is also shown in
Figure~\ref{fig:viral_load}. 

The measurement scale used in Figure~\ref{fig:immune-response} is logarithmic
and truncated.
In the adaptive immune response model the populations of virions and antibodies
are both represented as an integer, $n$. 
This integer $n$ is interpreted as the logarithm of the size of the population.
The justification for this  interpretation is the nature of testing with PCR and
laboratory assays.
Quantitative results are obtained by either successive dilutions (titrations)
or by successively culturing under conditions that permit
exponential growth.
The cycle threshold value reported for PCR tests, for example, is the number of
growth cycles required to detect RNA from the virus, and can be interpreted
(with a change of sign, up to a constant factor) as the logarithm of the initial
amount of RNA.
Similar reasoning applies to the titres reported from laboratory assays for
antibodies or antigens.
In both cases, a positive or negative result is simply a statement that a
threshold value has been passed.
The measurement scale $n$ that we use is truncated for practical reasons at a
maximum value $n_{\text{max}}$.
The rates at which various processes in the model occur are expressed in terms
proportional to $n$ for the various entities involved. 
This interpretation, together with the fitting procedure as described rests on 
the following assumption connecting the mechanistic model to the empirical test data:
\emph{the rate of positive test results is proportional to the logarithm
of the viral population.}

\subsection{Viral load distribution}
\newcommand{\tsbar}[2]{
  \begin{tikzpicture}
    \pgfplotsset{every y tick label/.append style={font=\footnotesize}}
    \begin{axis}[
      ybar,
      width=0.8\textwidth, height=0.125\textheight,
      xmin=0, xmax=20, ymin=0, ymax=0.6,
      no markers,
      xticklabels=none,
      ]
      \addplot+[ybar,bar width=10pt] table[x={c}, y={t#2}] {#1};
      \node at (axis cs:18,0.3) { $t = #2$};
    \end{axis}
  \end{tikzpicture}
}
\newcommand{\vlbar}[1]{\tsbar{viral-load.tsv}{#1}}
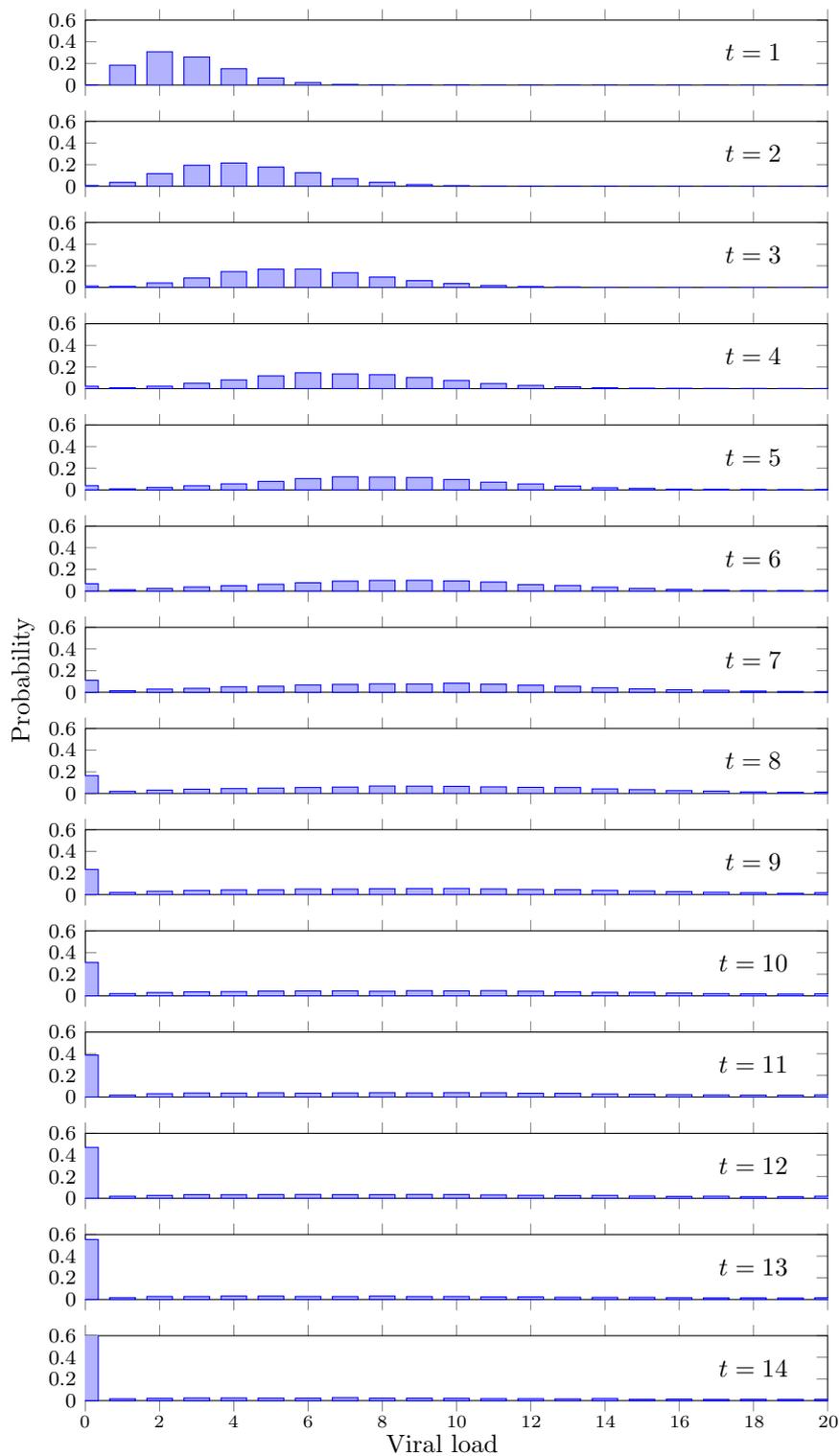
\begin{figure}
  \begin{center}
    \begin{tikzpicture}
      \foreach \n in {1,2,3,4,5,6,7,8,9,10,11,12,13,14} {
        \pgfmathsetmacro\y{-1.4*\n}
        \node at (0,\y) { \vlbar{\n} };
      }
      \foreach \v in {0,2,4,6,8,10,12,14,16,18,20} {
        \pgfmathsetmacro\x{-4.9+0.51*\v}
        \node at (\x,-20.25) { \scriptsize $\v$ };
      }
      \node at (0,-20.55) { Viral load };
      \node[rotate=90] at (-5.75,-10) { Probability };
    \end{tikzpicture}
  \end{center}
  \caption{Timeseries of probability distributions of having a viral load on a
scale of 0 to $n_{\text{max}}=20$, from the same simulation of 10,000
individuals as Figure~\ref{fig:immune-response}. By day 14, $80\%$ of viral load
is concentrated in approximately $20\%$ of individuals.}
  \label{fig:viral-load-timeseries}
\end{figure}
Figure~\ref{fig:viral-load-timeseries} shows data from the same simulation of
10,000 individuals. All individuals are infected at the beginning of the simulation. It depicts the probability distributions of an individual to
have a particular viral load as measured on the scale that we have defined, from
0 to $n_{\text{max}} = 20$.
We observe a wave of probability starting with a certainty that all individuals
have a viral load $n=1$, with viral load increasing as time progresses.
After about 10 days, a significant result appears: a long-tailed distribution
where an ever-smaller proportion of the population has a large viral load.
This is significant because it suggests a biological basis for the phenomenon of
over-dispersion in infectiousness, typically captured in infectious disease models by
asserting an appropriately parametrised Gamma distribution of
infectiousness~\cite{endo_estimating_2020}.
By the end of this 14 day sequence, $80\%$ of viral load is concentrated in
approximately $20\%$ of the population and by day 21 this has shrunk to $3\%$.
On average, over the entire 30 day period of the simulation, $80\%$ of the viral
load is concentrated in $21\%$ of the population.
Teasing out the implications of these results can shed some light on the phenomenon of superspreading individuals. On the one hand, increasing viral load may increase the chance of an individual transmitting virus in any given encounter. On the other hand, increasing viral load may also increase the chance of developing symptoms, to the extent that the individual would be more likely to self-isolate. Thus the chance of encounters occurring would diminish. The product of these two factors gives the overall chance of transmission. The interplay between these two factors: transmissibility upon a given encounter (increasing with viral load) and chance of encounter (decreasing with viral load) may result in the overall chance of transmission being a concave function of viral load, with a peak at an intermediate viral load. Factors that increase the chance of encounter at any given viral load (such as lack of paid sick leave from a public-facing job) would shift the peak of the function right (to a higher viral load).
However, if an individual with a large viral load is present in a situation
where transmission is likely, then it is reasonable to expect that a
superspreading event may occur~\cite{goyal_viral_2021}.

In passing, we observe that the processes in the adaptive immune response model
are more than are strictly required to reproduce this kind of long
tailed distribution of viral load.
It is possible to show that a simpler model where load simply increases
incrementally and then decreases, with the turning point chosen stochastically
will produce a similar result.
Such a simple model, however, lacks biological realism and fails to provide a causal or
mechanistic account of the phenomenon.

\section{Transmission dynamics}
\label{sec:transmission}
Unlike immune response which occurs within individuals, transmission is a
population-level process that occurs between individuals.
With the above formulation in hand, a simple model of transmission is very
natural,
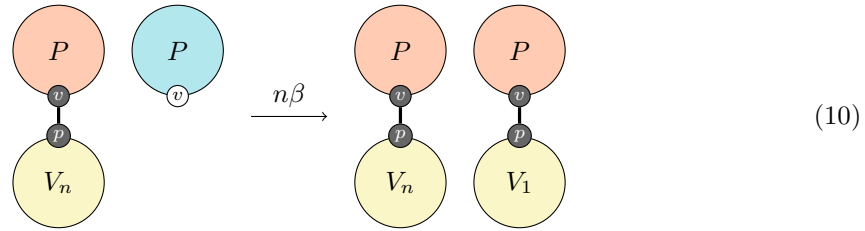
\begin{equation}
  \label{eq:transmission}
  \infectionl \arate{n\beta} \infectionr
\end{equation}
This says that an individual with some virus encounters an individual without
the virus (this implies well-mixed, mass action kinetics) and as a result of
that interaction, the second individual gets a small virus population.

Note in particular that it is not necessary to speak of any previous infections experienced by the second individual.
Nothing corresponding to immunity is required here because \emph{if} an
individual with established immune memory is exposed to the virus, \emph{then}
their immune response will simply be very rapid and they will clear the virus
quickly.
There is no need to impose an \emph{a priori} restriction on individuals being
exposed more than once or to assert a sharp distinction between individuals that
are ``immune'' and those that are not.

\begin{figure}
  \begin{subfigure}{\textwidth}
    \begin{tikzpicture}
      \begin{axis}[
        cycle list name=lshtm,
        width=\textwidth, height=0.25\textheight,
        xmin=0, xmax=90, ymin=0, ymax=1,
        xlabel={Time (days)}, ylabel={Fraction of population},
        no markers
        ]
        \envelope{S}{Sstd}{epidemic.tsv}\addlegendentry{S};
        \envelope{I}{Istd}{epidemic.tsv}\addlegendentry{I};
        \envelope{R}{Rstd}{epidemic.tsv}\addlegendentry{R};
        \draw[thick,dashed] (axis cs:30,0) -- (axis cs:30,0.23) -- (axis cs:56,0.23) -- (axis cs:56,0);
        \node at (axis cs:10,0.34) { $I(30) \approx I(56)$ };
        \path[draw] (axis cs:18,0.34) edge[out=0,in=130,-latex] (axis cs:30,0.24);
        \path[draw] (axis cs:18,0.34) edge[out=0,in=150,-latex] (axis cs:55,0.24);
        \draw[thick,dashed] (axis cs:20,0) -- (axis cs:20,0.04) -- (axis cs:66,0.04) -- (axis cs:66,0);
        \node at (axis cs:8,0.17) { $I(20) \approx I(66)$ };
        \path[draw] (axis cs:16,0.17) edge[out=0,in=140,-latex] (axis cs:20,0.05);
        \path[draw] (axis cs:16,0.17) edge[out=0,in=0] (axis cs:45,0.17);
        \path[draw] (axis cs:45,0.17) edge[out=0,in=150,-latex] (axis cs:66,0.05);
      \end{axis}
    \end{tikzpicture}
    \caption{Epidemic curve showing the susceptible ($S$), infectious ($I$)
      and removed ($R$) observables for a population of 10000 individuals
      calibrated for a reproduction number of 3. Envelopes show one and two
      standard deviations over 128 simulations. Marked on the graph are two
      pairs of time points where the mean number of infectious individuals are
      equal as the epidemic rises and falls.}
    \label{fig:epicurve}
  \end{subfigure}
  \begin{subfigure}{\textwidth}
    \begin{tikzpicture}
      \begin{axis}[
        ybar=0pt, cycle list name=lshtm,
        width=\textwidth, height=0.25\textheight,
        xmin=-0.5, xmax=20.5, ymin=0, 
        no markers, xlabel={Viral load}, ylabel={Fraction of infectious population},
        ]
        \addplot+[ybar,bar width=0.1,fill=lshtm1!40] table[x expr={\thisrow{c}}, y={t20}] {dyn-viral-load.tsv};\addlegendentry{$t = 20$};
        \addplot+[ybar,bar width=0.1,fill=lshtm2!40] table[x expr={\thisrow{c}}, y={t30}] {dyn-viral-load.tsv};\addlegendentry{$t = 30$};
        \addplot+[ybar,bar width=0.1,fill=lshtm3!40] table[x expr={\thisrow{c}}, y={t42}] {dyn-viral-load.tsv};\addlegendentry{$t = 42$};
        \addplot+[ybar,bar width=0.1,fill=lshtm4!40] table[x expr={\thisrow{c}}, y={t56}] {dyn-viral-load.tsv};\addlegendentry{$t = 56$};
        \addplot+[ybar,bar width=0.1,fill=lshtm5!40] table[x expr={\thisrow{c}}, y={t66}] {dyn-viral-load.tsv};\addlegendentry{$t = 66$};
      \end{axis}
    \end{tikzpicture}
    \caption{Viral load distribution at different points of the epidemic
      trajectory showing a rising $t \in \{20,30\}$, stationary $t = 42$, and
      falling $t \in \{56,66\}$ epidemic. Viral load in arbitrary logarithmic
      units. The probability masses of distributions are shifted to the left (lower viral loads)
      for a rising epidemic and the distribution for a falling epidemic is in fact
      bimodal with most infected individuals on the point of recovery but a
      significant number with slowly decaying high viral loads.}
    \label{fig:epiload}
  \end{subfigure}
  \caption{Epidemic curve and viral load distributions for a rising, stationary
    and falling epidemic.}
\end{figure}
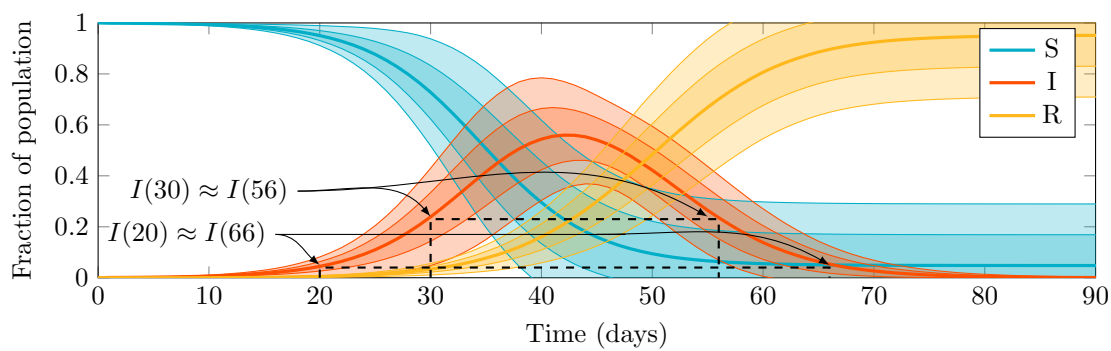
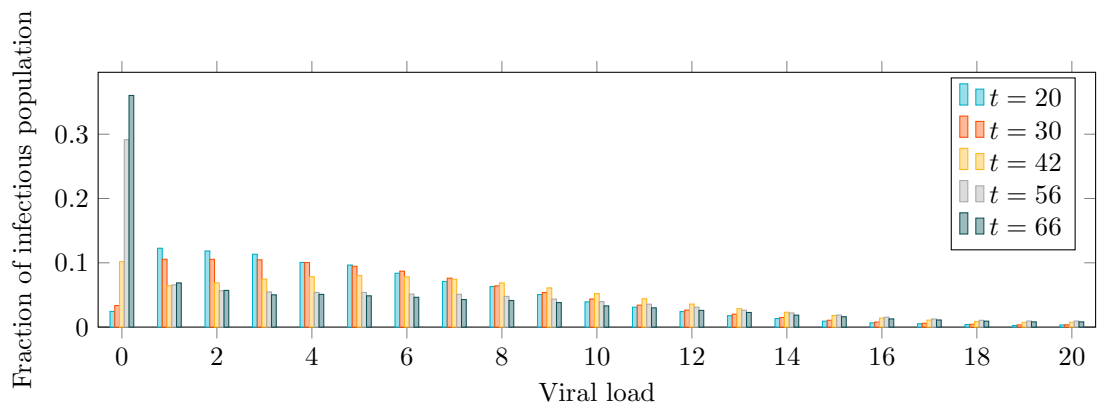

Figure~\ref{fig:epicurve} shows the epidemic curve produced by the combined
transmission and immune model according to the observables (cf
Section~\ref{sec:immune}\ref{sec:observables}) corresponding to the standard
epidemic model compartments.
The rate parameter $\beta$ in Equation~\ref{eq:transmission} is fit so as to
produce an epidemic with a basic reproduction number $R_0 = 3$, typical of the
ancestral strain of SARS-CoV-2 in a Western European setting.
Determining a precise value for $R_0$ and hence $\beta$ is setting-specific,
strongly affected by behaviour, and of only peripheral interest here.
Two pairs of time points are marked on the figure, corresponding to the times at
which, on average, the same number of individuals are infected as the epidemic
rises and falls.
The first pair, $t \in \{20,66\}$, is when a small number of individuals are
infected and the second, $t \in \{30,56\}$ when a much larger number are.
Finally, we notice that the epidemic is stationary (i.e. peaks, is neither
rising nor falling) around $t=42$.

Let us look at the viral load distributions at these chosen time points in more
detail, Figure~\ref{fig:epiload}.
We see that the distributions differ substantially.
For a rising epidemic ($t \in \{20,30\}$), the bulk of the probability mass is
shifted towards the left, indicating a lower viral load.
As individuals become infected faster and faster, there are more individuals who
are at an early stage in the course of their infection.
Recall from Figure~\ref{fig:viral_load} that we expect a peak viral load perhaps
a week after initial infection and then a slow decay over the subsequent several
weeks.
On the other hand, for a falling epidemic ($t \in \{56,66\}$), the probability
distribution is bimodal.
Much of the probability mass is concentrated in the lowest viral load
as many individuals are on the point of recovery.
However, a significant amount of probability mass is present in high viral loads
reflecting individuals with developed infections and slowly decaying viral loads.
A similar effect was noted by Hay and
colleagues~\cite{hay_estimating_2021} who showed that rising and falling
epidemics can be distinguished by looking at the distribution of cycle threshold
values in PCR tests and that this can, in fact, be used to estimate the
time-varying reproduction number $R(t)$ in a population.
We recover this result as a consequence of the immune dynamics in the model.

\section{Discussion}

Previous work on models, both across domains and for infectious disease modelling specifically~\cite{manheim_improving_2016}, have suggested that there are various features which make models more useful for different tasks. 
For infectious disease modelling, useful features include predictive accuracy,
ability to effectively model historic behaviour and predict the outcomes of future interventions in a way that
reveals causality, flexibility to update both the model structure
and calibration of parameters as new information becomes available, computational
tractability, and ease of explanation to policymakers and the public.
The approach presented here has significant
advantages across many of these desiderata though there is also still work to be done.
Our emphasis on \emph{modularity} and \emph{compositionality} facilitates the
flexibility to update model structure: individual rules or entire submodels can
be changed or substituted locally, without requiring global changes.
Indeed it is possible to analyse the dependencies between rules to determine
precisely the extent to which modifications to one part of the model may require
modification to other parts.
The modular structure means that it is often possible to calibrate
submodels individually, saving the great computational expense of performing this
operation for all parameters, which is often necessary for monolithic models.

Additionally, this modular and compositional approach facilitates
multidisciplinary collaboration. 
Infectious disease modelling, immunological modelling, modelling of behaviour
and economies and supply chains all require a substantial amount
of domain expertise and each has a substantial literature of sophisticated
models particular to that field.
By encapsulating the models and paying attention to the (often unclear or fuzzy) boundaries between them,
there is hope for constructing larger, richer models without the tight coupling
required by monolithic approaches.
Not only that, but we can mix and match.
Perhaps a particular immune model is computationally intensive and the cost
outweighs the benefit for the question at hand; it can simply be replaced with a
simpler one with a compatible interface.
Quantitative models of behaviour are difficult to calibrate because data to
relate them to ground truth are scarce.
This becomes a surmountable obstacle because one can simply use several; such
models are simply assumptions of dynamics about which our knowledge is limited
so we can evaluate the entire system under different sets of assumptions ranging
from simple to very sophisticated. The ability to slot these different models into the same context, can allow us to take advantage of decision-making frameworks that incorporate multiple predictors, each of which individually may be ``weak'' (i.e., low-confidence). \emph{Boosting} is an example of such an ensemble method~\cite{Schapire90thestrength} \cite{FREUND1995256}.
The structure described is that of \emph{Open
Systems}, a concept that dates at least to von Bertalanffy's work in
1950~\cite{von_bertalanffy_theory_1950}.
Recent advances in mathematics for interacting systems~\cite{fiadeiro_structured_2007}
stochastic rewriting systems~\cite{behr_nicolas_rule_2020} (of which the models
presented in this paper are examples) and specific kinds of open systems such as Petri
nets~\cite{baez_open_2020,baez_structured_2020}, and economic
games~\cite{ghani_compositional_2018,hedges_morphisms_2018} 
are placing these ideas on solid theoretical ground for computation.
By leveraging these theoretical advances we stand to make great practical
benefit in multi-scale and multi-disciplinary modelling. Even more ambitiously, such approaches allow representations used in sets of models to scale from physics to chemistry to biology to epidemiology, pharmacology, and more. This is consistent with similar approaches being taken in systems biology~\cite{Mirschel2009}, pharmacology~\cite{Hunt2009}, and (bio)chemical engineering~\cite{logist2011robust}. Of course, any one model or set of models will occupy only a small part of this spectrum; however, through composition and selection of models, many types of interdisciplinary and trans-disciplinary modelling approaches become possible, enabling us to address an immense variety of questions. 

There are numerous challenges to fully realising this dream.
A major obstacle is that there is, at present, no developed theory of how the
behaviour of one model influences the parameters of others - though some work in other fields is beginning to address this~\cite{Nimmegeers2016}.
This challenge is reflected in our discourse above where we explicitly choose how rates
(of transmission, say) depend on values (e.g. viral load) that are produced as a
result of a different model.
This pattern, where the state of the system as described by one process
influences the parameters of a different process, is very common and, at
present, must simply be handled manually.
A promising approach is that of Open
Games~\cite{ghani_hedges_winschel,ghani_compositional_2018,hedges_morphisms_2018} where the
composition of games is defined for \emph{all possible} utility functions --
essentially parametrisation of the games.
We could imagine Petri nets with rates or more general kinds of stochastic
rewriting systems taking a similar approach where their composition defined for all possible rate
functions with specialisation to particular choices of rate function done
post-hoc once the complete model is assembled.
The combination of Open Games and Open Stochastic Rewriting Systems is potentially very powerful.
Game theory is a unifying framework that can be used to underpin statistics~\cite{Wald1951StatisticalDF}, machine learning~\cite{SchapireFreundGame} \cite{cesa2006prediction}, microeconomic theory~\cite{kreps1990course} and the behavioural sciences~\cite{gintis2014bounds}, and is already used in a more specialised way in epidemiological modelling~\cite{reluga2021}. The Open Games formalism for compositional modelling supports combining submodels that may include games at different levels; for instance, within-host models where the agents are immune cells and virions within an organism, with between-host models where the agents are various organisms within a social milieu. 
This speculation suggests a path for incorporating epidemic models (which may themselves include game theory) into more overarching game-theoretic accounts in a principled way.

Another challenge is that the formulation of these models in the standard
$\kappa$ calculus permits only well-mixed interactions and site-graphs.
In other work~\cite{waites_transmission_2021}, we
have extended the language to support more general graphs of the kind that are
needed for epidemic models on networks, but a further extension to capture
explicit hierarchical notions of space as suggested in work on stochastic
bigraphs~\cite{krivine_stochastic_2008} would be beneficial.

The model of adaptive immune response given here is greatly simplified and we have provided no more than a vestigial model of the innate response. 
For COVID-19, it appears that the innate response plays an important role in whether an individual will recover after the initial acute phase of the disease or go on to develop severe disease~\cite{sette_adaptive_2021}.
The precise reason for this is not known. 
Equally, it is an open question why some individuals appear to recover completely and others develop persistent and varied symptoms known as ``long covid''~\cite{sudre_attributes_2021}. These persistent symptoms may be rooted in immune response or could have a physiological explanation.
In both cases, there is hope that more sophisticated models of immune response and physiology could provide the necessary insight.
Beyond these immediate questions, multi-scale, multi-system models are relevant not only for understanding population-level dynamics but for drug discovery and precision treatment of individuals.
Modular and compositional modelling techniques such as we demonstrate here provide a method for taming the substantial complexity involved.


\newcommand{\ethics}[1]{\paragraph{Ethics} #1}
\newcommand{\dataccess}[1]{\paragraph{Data Access} #1}
\newcommand{\aucontribute}[1]{\paragraph{Author Contributions} #1}
\newcommand{\funding}[1]{\paragraph{Funding} #1}
\newcommand{\ack}[1]{\paragraph{Acknowledgements} #1}
\newcommand{\disclaimer}[1]{\paragraph{Disclaimer} #1}
\newcommand{\competing}[1]{\paragraph{Competing Interests} #1}

\ethics{The ethical approval for the human data used in this analysis is detailed in the original manuscript reporting the study outcomes~\cite{houlihan_pandemic_2020}. It states that: ‘The study protocol was approved by the NHS Health Research Authority (ref 20/SC/0147) on 26 March 2020. Ethical oversight was provided by the South Central Berkshire Research Ethics Committee’.}
\dataccess{Source code is available at:\\\hfill\url{https://git.sr.ht/~wwaites/immune-transmission}}\\
Calibration data and sampled trajectories are available at:\\
\hfill{\url{https://datashare.ed.ac.uk/handle/10283/4056}}
\aucontribute{WW wrote the software, conducted the simulations, and wrote the
  initial draft of this article. WW, MC, VD and TH designed the numerical
  experiments. TR contributed empirical data. All authors
  wrote and revised the manuscript.}
\competing{The authors declare no competing interests}
\funding{WW, DM and TH acknowledge support from the Foundation for Innovative
  New Diagnostics. WW and RME were supported by MRC grant MR/V027956/1. DM also acknowledges support from the Center for Effective Altruism's Long-Term Future Fund.}
\ack{The authors would like to thank members of the CMMID COVID-19 Working Group
  for constructive feedback on drafts of this article.
  Simulations performed using resources provided by the Cambridge Service
  for Data Driven Discovery (CSD3) operated by the University of Cambridge
  Research Computing Service.}
\disclaimer{The funders had no role in the design, conduct or analysis of the
  study or the decision to publish.} 

\small
\bibliographystyle{plainnat}
\bibliography{id}

\normalsize
\appendix
\section{The combined model}
This model contains all of the rules that are described above, in
machine-readable form.

\label{sec:code}
\begin{lstlisting}[language=kappa]
// -*- mode: kappa -*-

// Rule-based model of a multi-scale epidemic consisting of an adaptive
// immune response and transmission
//
// Copyright 2021 William Waites <william.waites@lshtm.ac.uk>
//
// This program is free software: you can redistribute it and/or modify
// it under the terms of the GNU General Public License as published by
// the Free Software Foundation, either version 3 of the License, or
// (at your option) any later version.
//
// This program is distributed in the hope that it will be useful,
// but WITHOUT ANY WARRANTY; without even the implied warranty of
// MERCHANTABILITY or FITNESS FOR A PARTICULAR PURPOSE. See the
// GNU General Public License for more details.
//
// You should have received a copy of the GNU General Public License
// along with this program.  If not, see <https://www.gnu.org/licenses/>.

// This program is written in the Stratified Kappa language which is
// essentially a version of Kappa as understood by the KaSim simulator
// with a pre-processor that allows generation of rules from fragments
// of python code.

// exec
## res is passed in as a parameter and controls the resolution
## of counters in the simulation, and headroom to obviate truncation.
## typical values are 10 and 10.
## need to coerce to an integer because they may have been passed
## in as a floating point number.
res = int(res)
head = int(headroom)
// end exec

%agent:  Virus(person count{=0 / += $res+head$})
%agent:  Person(bcell, virus)
%agent:  BCell(person antibody affinity{=0 / += $res$})
%agent:  Antibody(bcell count{=0 / += $res+head$})

// Eq. 3.1 -- virus proliferation
// iterate for i in range(1,res+head):
'virus_$i$'  Virus(person[_], count{=$i$ / += 1})  @ k_virus/$res$
// end iterate

// Eq. 3.2 -- allocation of B-cell population
't_cell' Person(bcell[.], virus[_]), . \
      -> Person(bcell[1], virus[_]), BCell(person[1]) \
      @ inf

// Eq. 3.3 -- allocation of antibody population
'activate' Person(bcell[1], virus[_]), \
           BCell(person[1], antibody[.]), \
           .
        -> Person(bcell[1], virus[_]), \
           BCell(person[1], antibody[2]), \
           Antibody(bcell[2], count{=0}) \
        @ inf

// Eq. 3.4 -- affinity maturation
// iterate for i in range(res):
'mature_$i$' Person(bcell[1], virus[2]), \
             BCell(person[1], affinity{=$i$ / += 1}), \
             Virus(person[2], count{=v}) \
             @ v*k_mature/$res$
// end iterate

// Eq. 3.5 -- antibody production
// iterate for i in range(res+head):
'anti_$i$'  Person(bcell[1], virus[_]), \
            BCell(person[1], antibody[2], affinity{=a}), \
            Antibody(bcell[2], count{=$i$ / += 1}) \
            @ a*k_resp/$res$
// end iterate

// Eq. 3.6 -- virus neutralisation
// iterate for i in range(1,res+head+1):
'neutralise_$i$' Person(bcell[1], virus[2]), \
                 BCell(person[1], antibody[3]), \
                 Antibody(bcell[3], count{=c}), \
                 Virus(person[2], count{=$i$ / -= 1}) \
                 @ c*k_bind/$res$
// end iterate

// Eq. 3.7 -- clearing of the virus
'clear' Person(virus[1]), Virus(person[1], count{=0}) -> Person(virus[.]), . @ k_clear

// Eq. 3.8 -- waning
'waning'  Person(bcell[1], virus[.]), \
          BCell(person[1], antibody[2]), \
          Antibody(bcell[2], count{>=1 / -= 1}) @ k_wane

// Eq. 4.1 -- transmission
'infection' Person(virus[1]), Virus(person[1], count{=c}), \
            Person(virus[.]), . \
         -> Person(virus[1]), Virus(person[1]), \
            Person(virus[2]), Virus(person[2], count{=1}) \
         @ c*contacts*beta/N

// Initial conditions of a simulation -- rapidly allocate free virus populations to people
'start' Person(virus[.]), Virus(person[.]) -> Person(virus[1]), Virus(person[1]) @ inf

// Eq. 3.9 -- standard epidemiological observables
%obs: S |Person(bcell[.], virus[.])|
%obs: I |Person(virus[_])|
%obs: R |Person(bcell[_], virus[.])|

// Observables for various counters
// iterate for i in range(res+head+1):
%obs: V$i$ |Virus(person[_], count{=$i$})|
// end iterate
// iterate for i in range(res+head+1):
%obs: A$i$ |Antibody(count{=$i$})|
// end iterate
// iterate for i in range(res+1):
%obs: B$i$ |BCell(affinity{=$i$})|
// end iterate

// iniital conditions
%init: N   Person()
%init: I0  Virus(count{=1})
\end{lstlisting}
\end{document}